\documentclass[9pt,twocolumn]{extarticle}  

\usepackage[numbers,sort&compress]{natbib}
\usepackage[margin=0.7in]{geometry}  
\usepackage{times}
\usepackage{microtype}

\usepackage[compact]{titlesec}
\titlespacing*{\section}{0pt}{*1.5}{*0.5}
\titlespacing*{\subsection}{0pt}{*1.2}{*0.4}
\titlespacing*{\subsubsection}{0pt}{*1.0}{*0.3}

\usepackage{amsmath,amssymb,amsfonts}
\usepackage{mathtools}

\usepackage{graphicx}
\usepackage{booktabs}
\usepackage{multirow}
\usepackage{array}

\usepackage{algorithm}
\usepackage{algorithmic}

\usepackage{enumitem}
\setlist{nosep}  

\usepackage[colorlinks=true,
            linkcolor=black,
            citecolor=black,
            urlcolor=black]{hyperref}

\usepackage[numbers,sort&compress]{natbib}

\title{\textbf{Zero-Knowledge Proof Based Verifiable Inference of Models}}

\author{
  Yunxiao Wang \\
  \small Zhejiang University \\
  \small \texttt{wyx96922@gmail.com}
}

\date{} 

\begin{document}
\maketitle

\begin{abstract}

Recent advances in artificial intelligence (AI), particularly deep learning, have led to widespread adoption across various applications. 
Yet, a fundamental challenge persists: how can we verify the correctness of AI model inference when model owners cannot (or will not) reveal their parameters? These parameters represent enormous training costs and valuable intellectual property, making transparent verification difficult.
In this paper,
we introduce a zero-knowledge framework capable of verifying deep learning inference without exposing model internal parameters. Built on recursively composed zero-knowledge proofs and requiring no trusted setup, our framework supports both linear and nonlinear neural network layers, including matrix multiplication, normalization, softmax, and SiLU. Leveraging the Fiat–Shamir heuristic, we obtain a succinct non-interactive argument of knowledge (zkSNARK) with constant-size proofs.
To demonstrate the practicality of our approach, we translate the DeepSeek model into a fully SNARK-verifiable version named ZK-DeepSeek and show experimentally that our framework delivers both efficiency and flexibility in real-world AI verification workloads.

\end{abstract}


\section{Introduction}
\label{sec:intro}

As deep learning technologies advance at an unprecedented pace, artificial intelligence (AI) has shifted from laboratory prototypes to a central force shaping modern digital experiences. The release of ChatGPT-3.5 in 2022 powerfully illustrated this transformation, revealing just how naturally AI systems can reason, converse, and collaborate with humans. Yet beneath this remarkable progress lies a persistent challenge: the correctness of AI model inference remains fundamentally difficult to verify. In most practical deployments, AI models function as remote cloud services—users send API requests and receive outputs entirely controlled by the provider. These responses could be modified, simplified, or even generated by a different model altogether, and users would have no way of knowing. At the same time, model owners are understandably reluctant to reveal their parameters. Training large models demands massive computational investment, and the resulting weights represent valuable intellectual property at the core of an AI company’s competitive advantage.

This tension can be elegantly resolved using zero-knowledge proofs (ZKPs) \cite{goldreich1994definitions}. ZKPs make it possible to verify that a model's inference computation was executed correctly—without ever revealing the model parameters. In practice, an AI provider publishes a public commitment, such as a cryptographic hash uniquely representing the model’s parameters. When a user submits a request, the provider returns not only the inference result but also a corresponding zero-knowledge proof. This proof serves as a mathematical guarantee that the output was generated by the committed model, and the user can verify it efficiently using only the public commitment, learning nothing about the proprietary parameters themselves.

We present a zero-knowledge framework that enables deep learning models to be fully verifiable without revealing their internal parameters.Our construction is based on Kimchi \cite{o1Labs2023Kimchi}, a high-performance PLONKish proving system \cite{gabizon2019plonk}, and leverages universal and updatable structured reference strings (SRS), a practical requirement for real-world deployment. All neural network parameters, including matrix weights, RoPE coefficients, and softmax constants, are represented as integers to ensure compatibility with field arithmetic. The entire inference pipeline is decomposed layer by layer, and each layer into components such as general-purpose matrix multiplication or nonlinear transformations. When a component is too large to be handled efficiently within a single SNARK, it is recursively subdivided into rows and then into fixed-size segments. We generate proofs for these smaller pieces and merge them incrementally, ultimately producing a single proof attesting to the correctness of the entire inference process. Designed for scalability, the framework is capable of supporting extremely large models, including DeepSeek-V3 with its 671 billion parameters, while maintaining constant proof size and constant verification time.
 
Our implementation covers both linear and nonlinear neural-network operations, including general matrix multiplication, normalization, softmax, SiLU, cryptographic hash functions, Merkle tree constructions, and top-$k$ selection. Using the Fiat–Shamir heuristic \cite{bernhard2012not}, we instantiate a fully non-interactive zero-knowledge succinct argument of knowledge (zkSNARK). All components are realized through recursively composable SNARKs, enabling proofs to be efficiently merged and integrated into higher-level proofs.

To evaluate the framework, we construct ZK-DeepSeek, a SNARK-verifiable version of DeepSeek-V3 \cite{liu2024deepseek}. ZK-DeepSeek preserves the confidentiality of the model parameters while allowing anyone to independently verify its inference results. Our experiments show that the proposed framework achieves strong verifiability, modularity, and practical scalability for modern AI systems.

Our key contributions are summarized as follows.

\begin{itemize}

\item We propose an efficient framework for SNARK-based verification of neural-network inference, built on recursively composed proofs and requiring no trusted setup. The framework supports a wide variety of components, including matrix multiplication, normalization, and activation functions. Using the Fiat–Shamir heuristic, we obtain constant-size zkSNARK proofs, independent of model depth or architectural complexity.

\item We introduce a novel zero-knowledge proof construction for matrix multiplication between an input matrix $A \in \mathbb{Z}^{a \times n}$ and a weight matrix $W \in \mathbb{Z}^{n \times b}$, achieving a constraint complexity of $\mathcal{O}(an + nb)$. Once a proof for the weight matrix $W$ is generated, it can be reused in subsequent multiplications, significantly accelerating inference verification.

\item We implement ZK-DeepSeek, a SNARK-verifiable large-scale language model, to validate the practicality of our framework. We evaluate proof size, proving time, and verification time, demonstrating that our framework is both scalable and viable for real-world AI verification.

\end{itemize}



\section{Background and Related Work}
\label{sec:relwork}

As neural networks continue to advance, the value and impact of artificial intelligence (AI) have become widely recognized across both industry and academia. This growing momentum has attracted an increasing number of researchers and developers, driving the development of models that grow larger and more sophisticated each year. Yet, a fundamental question remains unresolved: how can we reliably verify the correctness of AI model inference?

Zero-Knowledge Proofs (ZKPs) and Trusted Execution Environments (TEEs) \cite{sabt2015trusted}  have emerged as two of the most promising approaches to this challenge. TEEs provide a secure hardware enclave in which sensitive computations can be executed in isolation, offering confidentiality and integrity. However, TEEs rely on proprietary hardware, suffer from limited memory, and remain vulnerable to various side-channel attacks, which limits their universality and scalability \cite{munoz2023survey}. ZKPs, on the other hand, allow a prover to demonstrate the correctness of a computation without revealing any underlying data or model parameters. Although ZKPs offer strong cryptographic guarantees, they have traditionally imposed high computational overhead and require complex circuit constructions. As cryptographic systems and computing hardware continue to mature, we expect these performance barriers to diminish.

Applying ZKPs to modern neural networks introduces its own set of challenges. First, large-scale models, particularly large language models (LLMs), may contain hundreds of billions of parameters, pushing current proof systems to their limits. Second, modern architectures rely extensively on nonlinear operations, which became central after the introduction of the transformer architecture in 2017 \cite{vaswani2017attention}. To address these nonlinearities efficiently, a variety of lookup-based techniques have been proposed in recent years, including Plookup \cite{gabizon2020plookup}, Caulk \cite{zapico2022caulk}, Caulk+ \cite{posen2022caulk+}, Flookup \cite{gabizon2022flookup}, Baloo \cite{zapico2022baloo}, and cq \cite{eagen2022cq}. These constructions represent a promising step toward scalable, efficient zero-knowledge verification for deep neural networks.

To tackle these challenges, we propose a SNARK-based framework for neural network verification requiring no trusted setup, built upon Kimchi, a modern and efficient PLONKish proving system. In this section, we first introduce PLONK and Kimchi, which form the foundation of our proof system. We then describe Pickles, the recursive composition framework that enables efficient proof chaining and incremental verifiable computation. Next, we discuss zkVC \cite{zhang2025zkvc}, whose matrix-multiplication techniques inspired our general-purpose multiplication scheme. Finally, we review several advanced architectural components of DeepSeek, demonstrating how our framework can seamlessly support and verify them within a zero-knowledge environment.

\subsection{Notations}

We denote the computational security parameter by $\lambda$. Throughout this work, we use 0-based indexing for all data structures. For integer ranges, we write $[a,b]$ to denote the set ${a, \ldots, b}$, and $[a,b)$ to denote ${a, \ldots, b-1}$.

Matrices and tensors are written in uppercase letters (e.g., $X$), vectors in lowercase letters (e.g., $v$), and individual elements using subscripts (e.g., $v_i$ for the $i$-th element of $v$, and $X_{i,j}$ for the entry in row $i$, column $j$ of $X$). We follow the PyTorch slicing convention:

\begin{itemize}
\item $v[a:b]$ extracts elements from index $a$ through $b-1$;
\item $X[a, c:d]$ selects row $a$ and columns $c$ through $d-1$;
\item $X[a:b, c:d]$ selects rows $a$ through $b-1$ and columns $c$ through $d-1$.
\end{itemize}

We use symbols of the form $Q_l$ to denote integers representing real values scaled by a factor of $2^l$. For a polynomial $P$, $deg(P)$ denotes its degree. We use $F_p$ to represent the finite field encoding unsigned integers in the range $[0, p-1]$, and $|F_p|$ to denote the size of the field.

\subsection{PLONK and Kimchi}

PLONK (Permutations over Lagrange-bases for Oecumenical Noninteractive arguments of Knowledge) is a universal and updatable zkSNARK protocol proposed by Gabizon et al. in 2019. Unlike earlier systems such as Groth16 \cite{groth2016size}, which require a circuit-specific trusted setup, PLONK introduces a universal structured reference string (SRS) that can be reused across arbitrary circuits. Its key innovations include a powerful permutation argument that efficiently enforces copy constraints across circuit wires, and the use of Lagrange-basis polynomials to encode arithmetic circuits directly over evaluation domains. Combined with KZG polynomial commitments \cite{kate2010constant}, PLONK achieves succinct proofs, fast verification, and a constant proof size. Its modular design has since inspired a family of enhanced systems (such as Turbo-PLONK \cite{gabizon2020proposal}, HyperPlonk \cite{chen2023hyperplonk}, and Kimchi) which further improve performance through lookup arguments, custom gates, and tighter recursive proofs.

Kimchi extends PLONK with a series of structural improvements, optimizations, and architectural refinements that significantly increase expressiveness and efficiency. Whereas PLONK’s constraint system is limited to 3 registers, Kimchi expands this to 12 registers grouped into two types: I/O registers, which support flexible connectivity across gates, and temporary registers, which are local to individual gates. This richer register set enables gates to operate on multiple inputs simultaneously, a capability that greatly simplifies the implementation of complex primitives. For example, scalar multiplication on elliptic curves naturally requires at least three inputs (a scalar and two coordinates of a curve point)—a task cumbersome to encode in standard PLONK but straightforward in Kimchi.

Recognizing that certain operations recur frequently in practical circuits, Kimchi also introduces a suite of specialized custom gates that implement them efficiently. These include gates optimized for scalar multiplication, the Poseidon hash function, and various encryption operations—nine new gates in total. Another important innovation is output forwarding, which allows a gate to write its output directly into the registers used by the next gate. This feature is particularly beneficial for iterative computations such as the Poseidon hash \cite{lorenzo2021poseidon}, where the same transformation must be applied repeatedly. Together, these enhancements make Kimchi significantly more flexible and efficient than classical PLONK, especially in recursive SNARK settings.

\subsection{Pickles}

Ben-Sasson et al. introduced the concept of cycles of elliptic curves  \cite{ben2017scalable}, in which two elliptic curves, commonly referred to as $Tick$ and $Tock$, are arranged so that each can efficiently verify proofs generated over the other. A proof produced on $Tick$ can be verified on $Tock$, and vice versa. This mutually verifiable relationship enables recursive proof composition: proofs can be repeatedly wrapped inside higher-level proofs, allowing arbitrarily complex computations to be verified succinctly through iterative proof chaining.

Pickles builds directly on this idea and serves as the inductive zkSNARK composition framework underlying the Mina Protocol \cite{bonneau2020mina}. It enables proofs to be generated, verified, and recursively composed in a highly modular and programmable way, supporting incremental verifiable computation across blockchain blocks and state transitions. Pickles adopts the Pasta curve pair (also known as $Tick$ and $Tock$) which alternate in their proving roles: $Tick$ handles the more complex proving circuits, while $Tock$ verifies the previous proof and produces a compact wrap proof. This alternating, two-curve architecture provides a scalable and efficient foundation for deep recursive proof trees, enabling constant-time verification even as the recursion depth grows. As a result, Pickles stands as one of the most mature and practical recursive SNARK systems deployed in production today.

\subsection{Constraint-Reduced Polynomial Circuits}

\begin{figure}
	\centering
		\includegraphics[trim={0cm 3cm 0cm 0cm},clip,scale=.2]{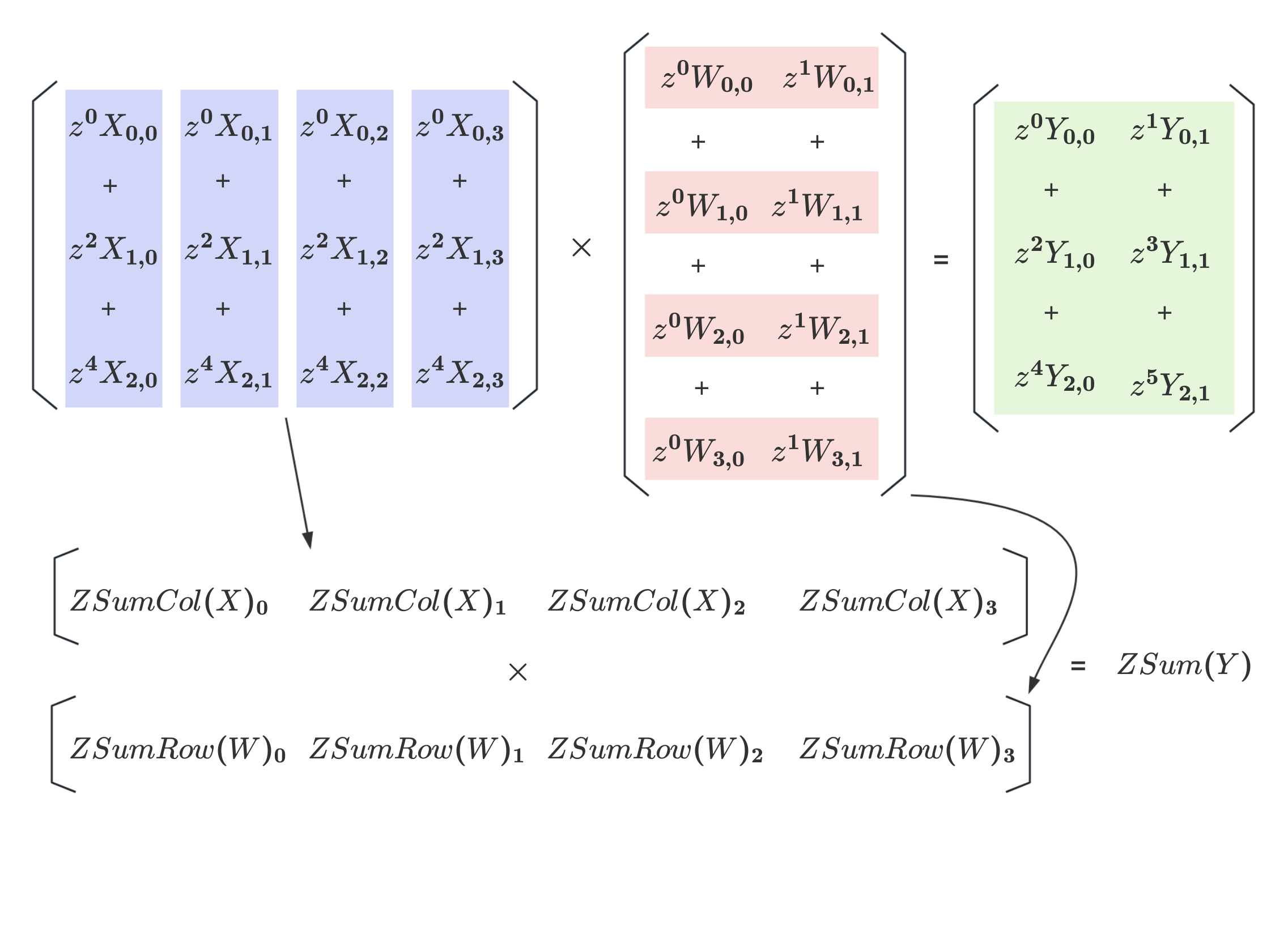}
	\caption{The multiplication in CRPC}
	\label{FIG:1}
\end{figure}

Constraint-Reduced Polynomial Circuits (CRPC), introduced in zkVC \cite{zhang2025zkvc}, aim to significantly reduce the cost of generating zero-knowledge proofs for matrix multiplication, one of the most computation-intensive operations in deep learning verification. As illustrated in Figure 1, CRPC maps each column of $X$ and each row of $W$ into univariate polynomials using a randomly chosen variable $z$, and expresses the output matrix $Y$ in the same manner. For example:

\begin{center}
$(z^0Y_{0,0} + z^1Y_{0,1} + z^2Y_{1,0} + z^3Y_{1,1} + z^4Y_{2,0} + z^5Y_{2,1})$

$= (z^0X_{0,0} + z^2X_{1,0} + z^4X_{2,0}) \cdot (z^0W_{0,0} + z^1W_{0,1})$

$+ (z^0X_{0,1} + z^2X_{1,1} + z^4X_{2,1}) \cdot (z^0W_{1,0} + z^1W_{1,1})$

$+ (z^0X_{0,2} + z^2X_{1,2} + z^4X_{2,2}) \cdot (z^0W_{2,0} + z^1W_{2,1})$

$+ (z^0X_{0,3} + z^2X_{1,3} + z^4X_{2,3}) \cdot (z^0W_{3,0} + z^1W_{3,1})$
\end{center}

This idea generalizes naturally to matrix multiplication $Y^{a \times b} = X^{a \times n} \times W^{n \times b}$:

\begin{equation}
    \sum_{i=0}^{a-1} \sum_{j=0}^{b-1} z^{ib+j} Y_{i,j} = \sum_{k=0}^{n-1} \left( \sum_{i=0}^{a-1} z^{ib} X_{i,k} \right) \left( \sum_{j=0}^{b-1} z^j W_{k,j} \right)
    \label{eq:placeholder_label}
\end{equation}

By the Schwartz–Zippel Lemma \cite{schwartz1980fast} \cite{zippel1979probabilistic}, this identity holds with overwhelming probability if and only if the matrix multiplication is correct. Based on this, we define:

\begin{equation}
ZMul(X) = \sum_{i=0}^{a-1} \sum_{j=0}^{b-1} z^{ib+j} X_{i,j}
\end{equation}

where $a$ and $b$ denote the height and width of $X$. We further define:

\begin{equation}
ZMulCol(X) = (\sum_{i=0}^{a-1} z^{ib} X_{i,k}))_{k = 0}^{n-1}
\end{equation}

\begin{equation}
ZMulRow(W) = (\sum_{j=0}^{b-1} z^j W_{k,j})_{k = 0}^{n-1}.
\end{equation}

CRPC effectively encodes the entire matrix multiplication into a single polynomial identity, reducing the constraint complexity from the naive $\mathcal{O}(anb)$ to $\mathcal{O}(an + nb)$. This reduction significantly decreases the number of arithmetic constraints, leading to much faster proving times in zkSNARK systems. As reported in zkVC, CRPC can reach complexity as low as $\mathcal{O}(n)$ when instantiated in the Groth16 proving system. In our PLONKish setting, the complexity becomes $\mathcal{O}(an + nb)$, still a major improvement over the original $\mathcal{O}(anb)$ complexity.

\subsection{DeepSeek}

\textbf{Multi-Head Latent Attention(MLA)}

DeepSeek introduces Multi-Head Latent Attention (MLA) to significantly reduce the memory and bandwidth overhead of attention mechanisms without compromising model quality. Rather than caching full per-head Key/Value (KV) tensors, MLA projects keys and values into a shared low-rank latent space and caches only these compact latent representations. During inference, head-specific KV features are reconstructed from the shared latents using lightweight projection layers. This design dramatically shrinks the KV cache, mitigates memory-bound throughput limitations, and substantially improves efficiency, especially for long-context inference, making large-scale deployment far more scalable. MLA integrates cleanly with standard architectural components (such as Rotary Positional Embedding) and aligns well with the memory layouts and optimized attention kernels used in production systems.

\noindent \textbf{Mixture-of-Experts(MoE)}

DeepSeek also employs a sparse Mixture-of-Experts (MoE) architecture, where a learned gating network selects a small top-$k$ subset of specialized feed-forward “experts” for each token. Because only a fraction of the total parameters are activated for any given token, MoE provides significantly higher model capacity at a much lower effective computational cost, improving scaling efficiency during both training and inference. DeepSeek’s MoE system emphasizes stable and balanced routing, communication-efficient expert parallelism, and load-balancing strategies that avoid harming the primary optimization objective. Combined with MLA, the MoE design yields an advantageous cost–performance profile: MLA reduces memory pressure in the attention module, while MoE improves model expressiveness and throughput. Together, they enable long-context, high-throughput, and deployment-friendly large language models.

\section{Methodology}
\label{sec:methodology}

We propose a comprehensive framework for constructing zero-knowledge proofs (ZKPs) tailored to neural network models. The framework includes all essential components of modern architectures, such as general-purpose matrix multiplication, embedding, normalization, and activation functions, enabling efficient and verifiable AI computation within a cryptographic setting.

In large neural networks, particularly large-scale language models (LLMs), both input tensors and weight matrices often reach thousands of dimensions. Encoding such high-dimensional data directly into a PLONK-style circuit causes a rapid increase in polynomial complexity, substantially increasing both proving and verification time. To address this scalability challenge, our framework partitions oversized matrices into smaller submatrices along both dimensions. Each submatrix is proven independently, and the resulting proofs are recursively merged to certify the correctness of the full computation.
Proofs from different matrices and components are likewise assembled incrementally, layer by layer, ultimately producing a single proof attesting to the correctness of the entire inference pipeline.
This recursive composition strategy enables the framework to scale gracefully to extremely large neural networks, ensuring practical verification without compromising soundness.

\subsection{General-purpose Matrix Multiplication (GeMM)}

For a matrix multiplication of the form $X^{a \times n} \times W^{n \times b} = Y^{a \times b}$ in neural network models, three essential correctness conditions must be verified:

\noindent \textbf{1. Input consistency:}
The input matrix $X$ must correspond exactly to the output produced by the preceding computation stage.

\noindent \textbf{2. Multiplicative correctness:}
The relation among $X$, $W$, and $Y$ must satisfy the polynomial identity associated with matrix multiplication, as defined in Equation (1).

\noindent \textbf{3. Model integrity:}
The weight matrix $W$ must match the committed model parameters(hashes), ensuring that the computation is performed on the intended neural network.

Constraint~1 is enforced by comparing $ZMul(X)$ with $ZMul(Y')$, where $Y'$ is the output of the previous component. By the Schwartz–Zippel Lemma, if $z$ is chosen randomly, the probability that $ZMul(X) = ZMul(Y')$ while $X \neq Y'$ is negligible.
Constraint~2 is verified using the identity
$$
ZMulCol(X) \cdot ZMulRow(W) = ZMul(Y),
$$
and Constraint~3 is enforced through the hash commitment of the weight matrix $W$.

To efficiently construct the proof, we proceed in three stages:

\textbf{1. Partitioning the input matrix $X$.}
The matrix $X$ is divided into small segments of size $1 \times s$ (zero-padded when $n$ is not divisible by $s$). For each segment, the prover computes $ZMul$ and $ZMulCol$ values to generate base proofs, which are later merged column-wise (see Figure~2).

\textbf{2. Partitioning the weight matrix $W$.}
Likewise, $W$ is split into segments of size $s \times 1$. Each segment’s prover computes $ZMulRow$ and the associated hash value, producing a set of base proofs that are merged row-wise (see Figure~3).

\textbf{3. Recursive proof composition.}
Each base proof from $X$ is paired with its corresponding base proof from $W$ to produce a combined proof for that fragment. These combined proofs are then recursively merged in a binary-tree structure, ultimately yielding a single proof attesting to the correctness of the entire matrix multiplication (see Figure~4).

\begin{figure}
	\centering
		\includegraphics[trim={0cm 5cm 0cm 0cm},clip,scale=.178]{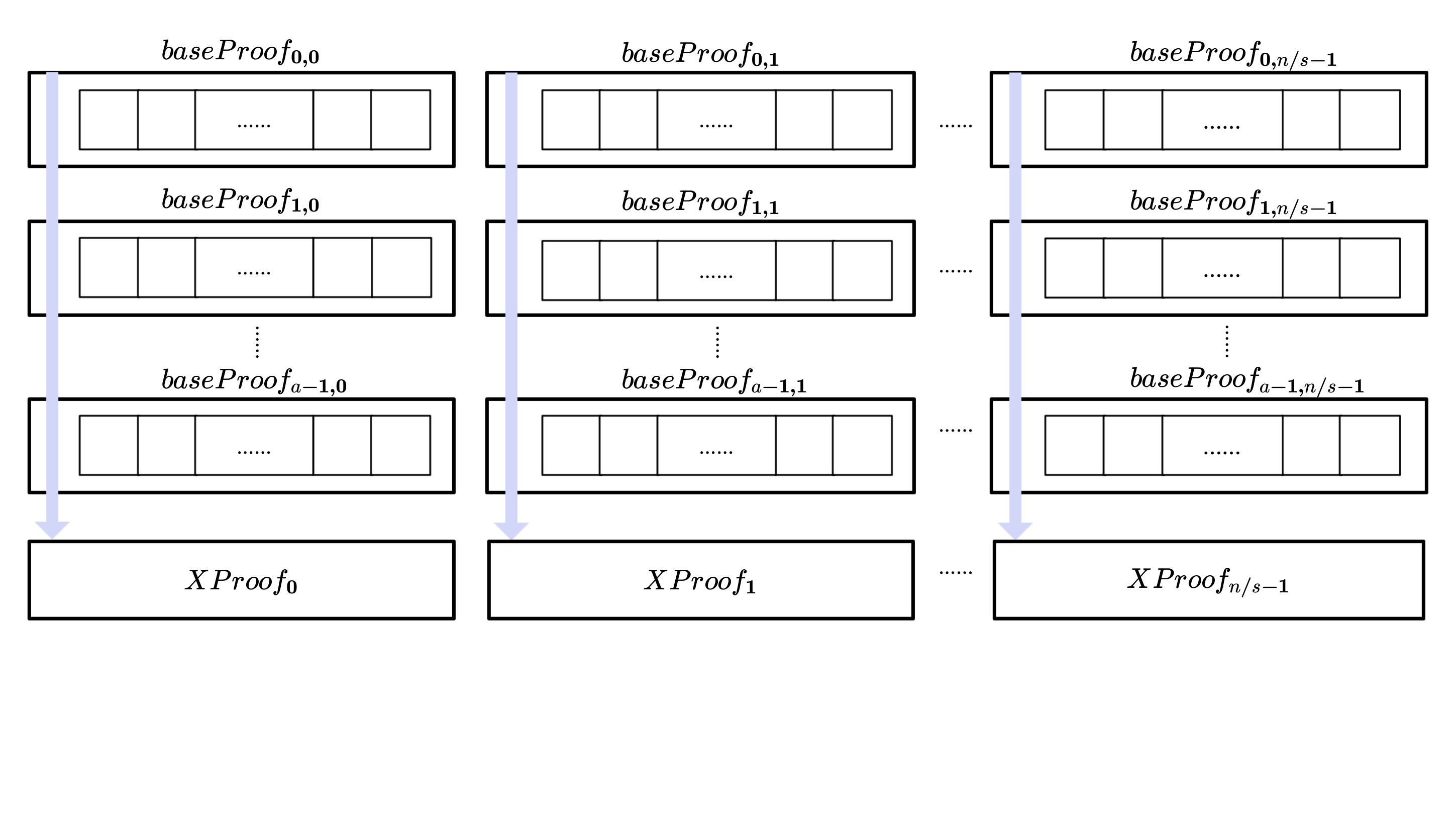}
	\caption{The proof construction of matrix X}
	\label{FIG:2}
\end{figure}

\begin{figure}
	\centering
		\includegraphics[trim={0cm 0cm 0cm 0cm},clip,scale=.22]{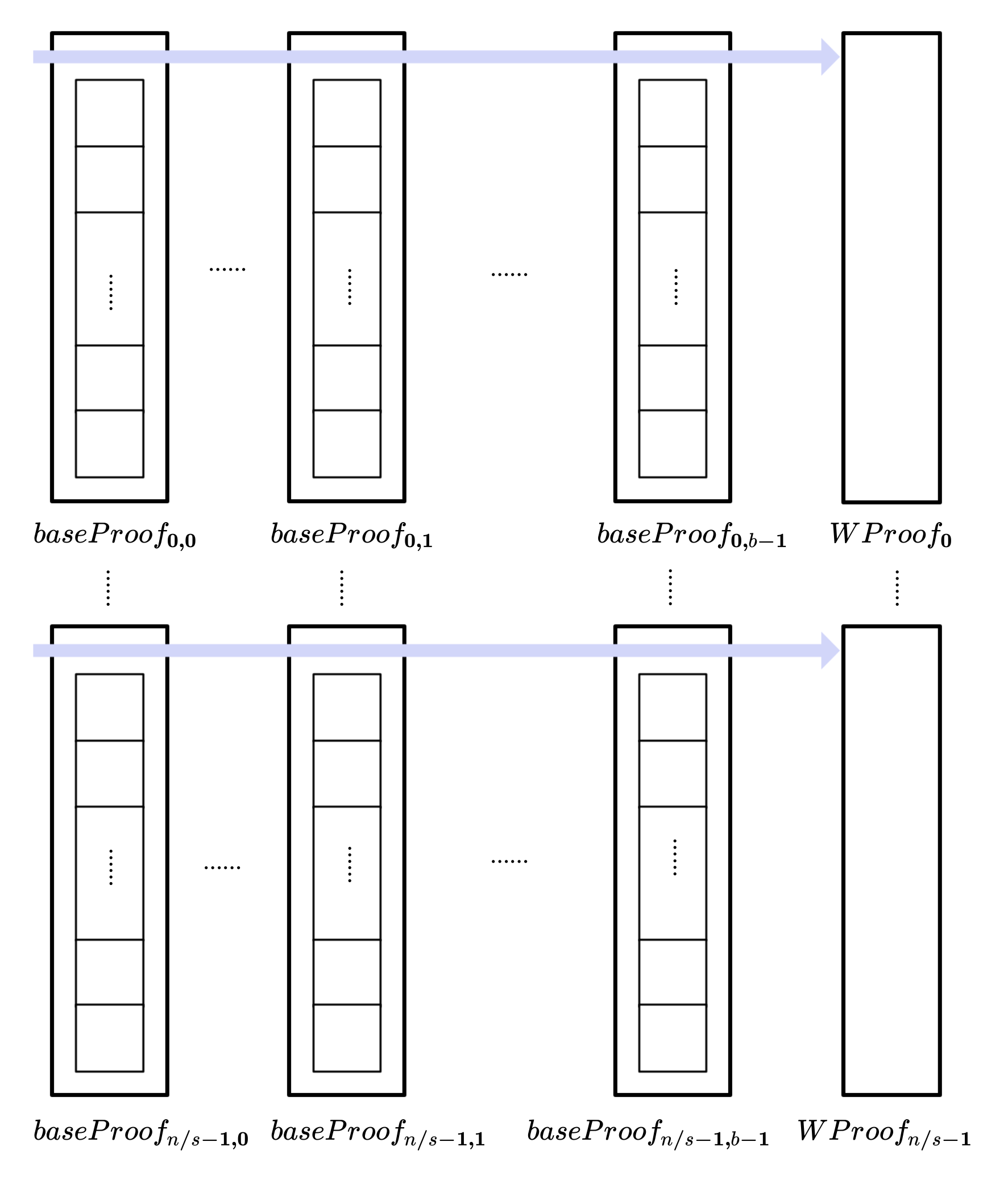}
	\caption{The proof construction of matrix W}
	\label{FIG:3}
\end{figure}

\begin{figure*}
	\centering
		\includegraphics[trim={0cm 3cm 0cm 0cm},clip,scale=.33]{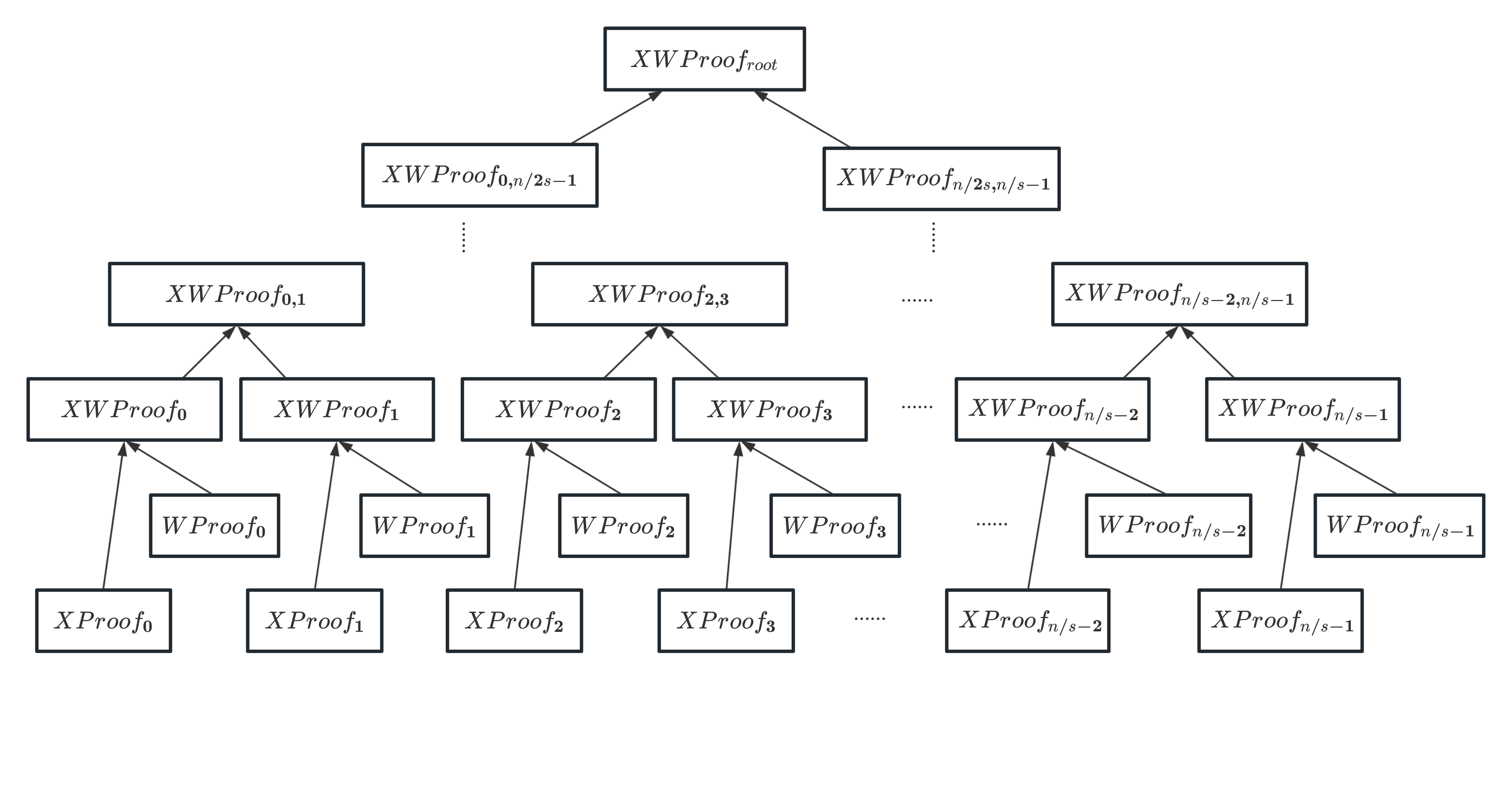}
	\caption{The proof composition of the multiplication of X and W}
	\label{FIG:4}
\end{figure*}

\subsection{Root Mean Square Layer Normalization (RMSNorm)}

Root Mean Square Layer Normalization (RMSNorm) \cite{zhang2019root} normalizes each input vector by dividing it by its root mean square (RMS), computed as:

\begin{equation}
    RMS(x) = \sqrt{\frac{1}{n}\sum_{1}^{n} x_i^2 + \epsilon  } 
\end{equation}

To make this operation compatible with zero-knowledge proof systems, our implementation reformulates the computation using pure integer arithmetic. As shown in Equation (5), both the averaging term $\frac{1}{n}$ and the square root function should be replaced with equivalent integer-based constraints. For numerical stability, we set $\epsilon = 1$. This leads to the following integer formulation:

\begin{equation}
\begin{split}
    RMS(x)^2 \leq Q + \epsilon < (RMS(x) + 1)^2 \\
    Q = \frac{{\textstyle \sum_{i=1}^{d} x_i^2} }{n}  \\
    R = ({\textstyle \sum_{i=1}^{d} x_i^2}) \% n \\
    {\textstyle \sum_{i=1}^{n} x_i^2} = Q \cdot n + R, R < n
\end{split}
\end{equation}

Given a matrix $X^{a \times n}$ and a weight vector $W^{1 \times n}$, each row of $X$ is normalized according to:

\begin{equation}
    (y_i)_{0}^{n-1} = (\frac{x_i \cdot w_i}{RMS(x)})_{0}^{n-1}
\end{equation}

To ensure correctness, three constraints are verified within the proof system.

1. The $ZMul$ value of $X$ must match the $ZMul(Y')$ value produced by the previous component.

2. The relationships in Equations (6) and (7) must be satisfied.

3. The hash of the weight vector $W$ must be computed and checked against the committed model parameters to ensure integrity.

4. The $ZMul$ value of $Y$ must be calculated to serve as input for the subsequent layer.

To efficiently construct the proof, we apply the following steps:

1. Both $X$ and $W$ are partitioned into segments of shape $1 \times s$ (with zero-padding when $n$ is not divisible by $s$). For each segment pair, the prover computes the segment’s $ZMul$ contribution and accumulates the squared-sum of $X$ (see Figure 5).

2. All segment proofs belonging to the same row are merged into a single proof that certifies the normalization of that row.

3. Row proofs are recursively merged to produce a final proof that attests to the correctness of the entire RMSNorm computation.

\begin{figure*}
	\centering
		\includegraphics[trim={0cm 1cm 1.5cm 0cm},clip,scale=.28]{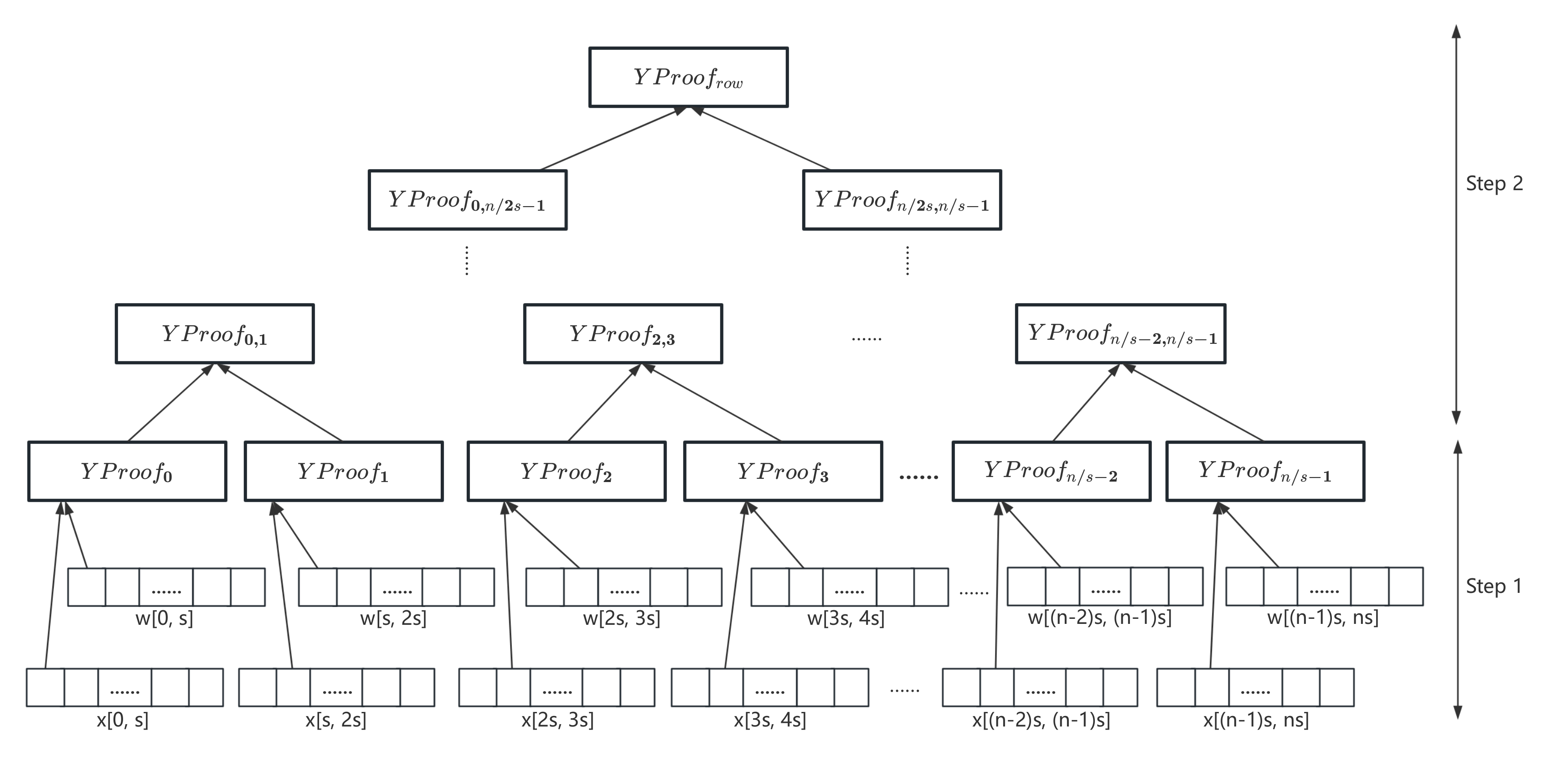}
	\caption{The row proof composition of the RMSNorm}
	\label{FIG:5}
\end{figure*}

\subsection{Embedding}

The embedding component transforms discrete input tokens into their corresponding vocabulary embeddings, establishing the foundation of the model’s representational space. In our framework, each vocabulary embedding is associated with a publicly known hash, allowing the verifier to confirm the integrity of the input provided to the circuit. The component’s primary function is therefore to compute both the hash and the $ZMul$ value of the input embedding matrix, ensuring that these two representations correspond to the same underlying input.

To certify correctness, the proof system verifies two conditions:

1. The hash of the input embedding matrix matches the value derived from the public vocabulary embedding hashes.

2. The $ZMul$ value of the embedding matrix is computed correctly so it can be used as input to the next layer.

To efficiently enforce these guarantees, we follow the procedure illustrated in Figure 6:

\textbf{1. Segmentation:} The input matrix $X$ is divided into smaller segments of size $1 \times s$. For each segment, the prover computes its hash and accumulates its contribution to the overall $ZMul$ value.

\textbf{2. Row-level aggregation:} The segment proofs are merged horizontally to form row-level proofs, each certifying the correctness of one row of embeddings.

\textbf{3. Embedding proof composition:} All row-level proofs are then recursively merged, producing a single proof that attests to the correctness and consistency of the entire embedding process.

\begin{figure}
	\centering
		\includegraphics[trim={0cm 6cm 0cm 0cm},clip,scale=.21]{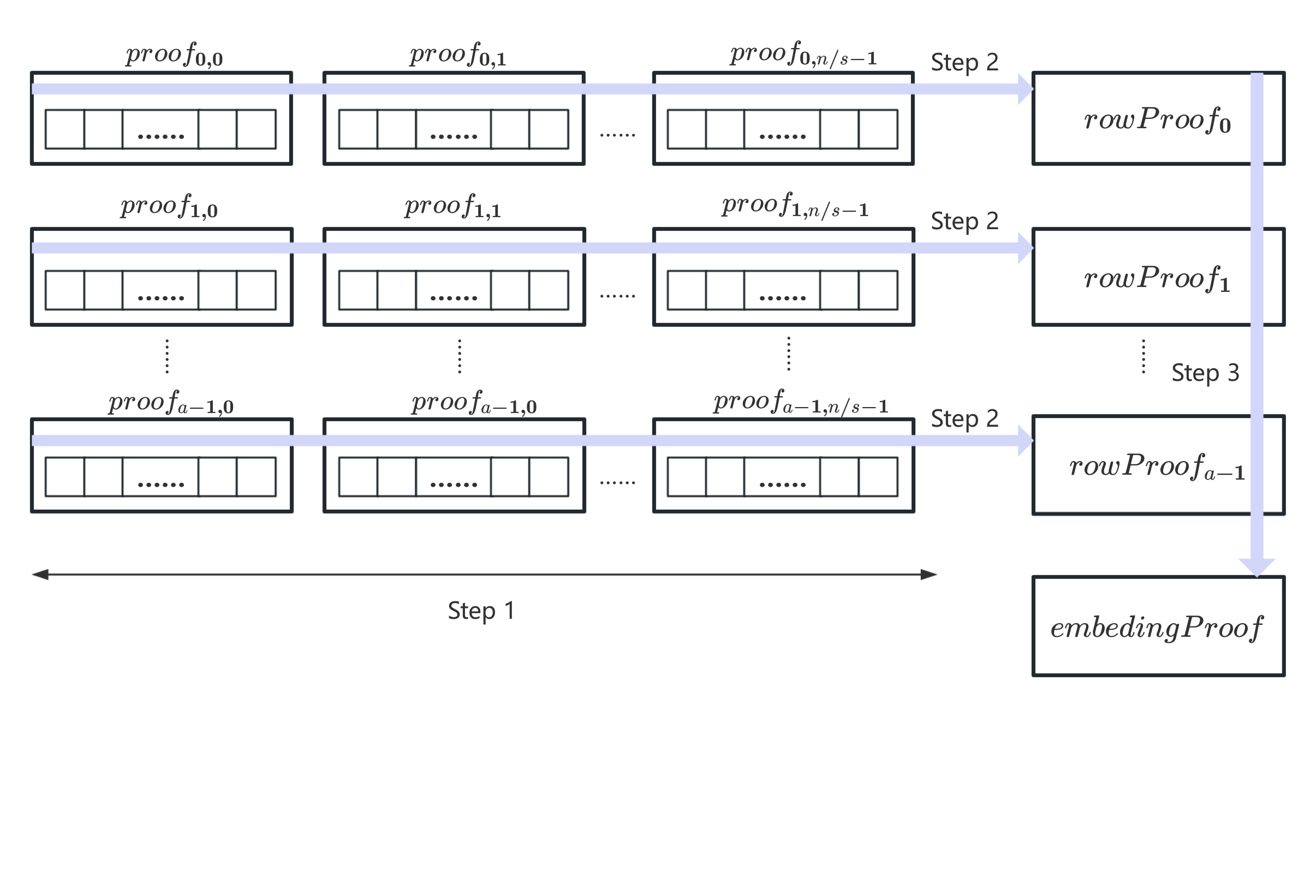}
	\caption{The proof composition of the embedding component}
	\label{FIG:6}
\end{figure}

\subsection{Rotary Positional Embedding (RoPE)}

Rotary Positional Embedding (RoPE) \cite{su2024roformer} encodes positional information by applying a dimension-wise rotation to the query and key vectors in each attention head, using paired sinusoidal functions. This rotation is expressed as:

{\scriptsize
\begin{equation}
RoPE\begin{pmatrix} q_{0} \\ q_{1} \\ q_{2} \\ q_{3} \\ \cdots \\ q_{d - 2} \\ q_{d - 1} \end{pmatrix} = \begin{pmatrix} q_{0} \\ q_{1} \\ q_{2} \\ q_{3} \\ \cdots \\ q_{d - 2} \\ q_{d - 1} \end{pmatrix} \otimes \begin{pmatrix} cos(m\theta_{0}) \\ cos(m\theta_{0}) \\ cos(m\theta_{1}) \\ cos(m\theta_{1}) \\ \cdots \\ cos(m\theta_{d / 2 - 1}) \\ cos(m\theta_{d / 2 - 1}) \end{pmatrix} + \begin{pmatrix} - q_{1} \\ q_{0} \\ - q_{3} \\ q_{2} \\ \cdots \\ - q_{d - 1} \\ q_{d - 2} \end{pmatrix} \otimes \begin{pmatrix} sin(m\theta_{0}) \\ sin(m\theta_{0}) \\ sin(m\theta_{1}) \\ sin(m\theta_{1}) \\ \cdots \\ sin(m\theta_{d / 2 - 1}) \\ sin(m\theta_{d / 2 - 1}) \end{pmatrix}
\end{equation}
}

To make RoPE efficient inside a zero-knowledge circuit, we precompute all sinusoidal weights($\cos(m\theta_i)$ and $\sin(m\theta_i)$) along with their hash commitments. These values are stored as a single weight matrix

{\footnotesize
\begin{equation}
\begin{pmatrix} cos(0\theta_{0}), sin(0\theta_{0}), cos(0\theta_{1}), \cdots cos(0\theta_{d/2-1}), sin(0\theta_{d/2-1}) \\ cos(1\theta_{0}), sin(1\theta_{0}), cos(1\theta_{1}), \cdots cos(1\theta_{d/2-1}), sin(1\theta_{d/2-1}) \\ cos(2\theta_{0}), sin(2\theta_{0}), cos(2\theta_{1}),  \cdots cos(2\theta_{d/2-1}), sin(2\theta_{d/2-1}) \\
\cdots \\cos(m\theta_{0}), sin(m\theta_{0}), cos(m\theta_{1}),  \cdots cos(m\theta_{d/2-1}), sin(m\theta_{d/2-1})\end{pmatrix},
\end{equation}
}

\noindent allowing the SNARK prover to perform only arithmetic operations against this matrix while verifying hash consistency. The architecture for RoPE proof generation across multiple attention heads is shown in Figure 7.

To efficiently produce a SNARK proof for the full RoPE computation, we follow the workflow below:

\begin{figure}
	\centering
		\includegraphics[trim={0cm 6cm 0cm 0cm},clip,scale=.2]{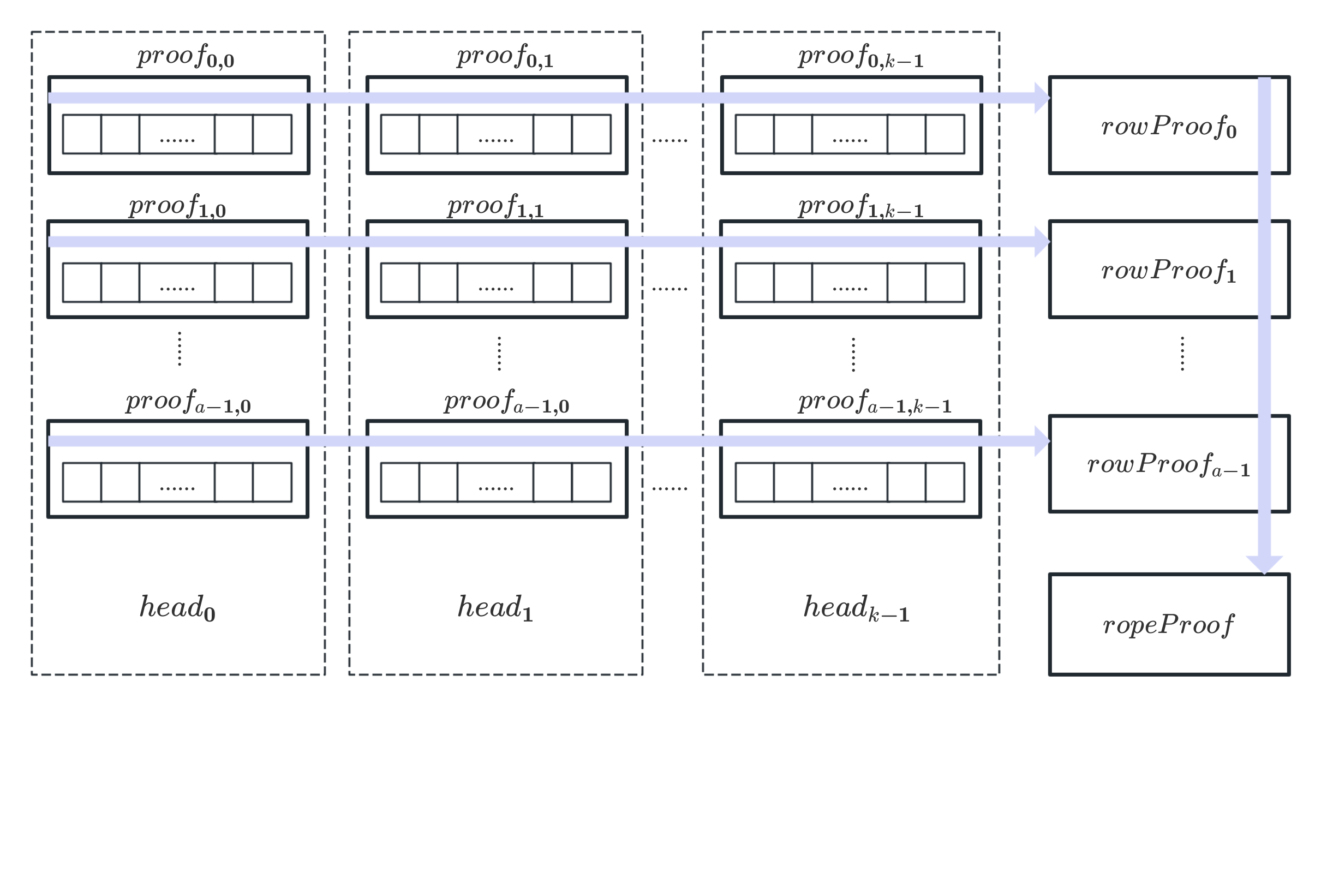}
	\caption{The proof composition of the RoPE component}
	\label{FIG:7}
\end{figure}

\textbf{Per-head computation:}
The input matrix $X$ is divided into $m$ rows, each containing $k$ attention heads. For every head, the prover computes

$$Y_{i, head} = X_{i,head}W_i, $$
where $W_i$ is the $i$-th RoPE weight row. During this step, we accumulate the $ZMul$ values of both $X_{i,\text{head}}$ and $Y_{i,\text{head}}$, as well as the hash of $W_i$.

\textbf{Row-level aggregation:}
Head-level proofs are merged horizontally to form row-level proofs, each certifying the correctness of one full row of RoPE computation.

\textbf{Row wrapping:}
Each row proof is wrapped into a new proof that additionally verifies that the RoPE weight hashes match the expected commitments.

\textbf{Global merging:}
Finally, all wrapped proofs are recursively merged to produce a single proof attesting to the correctness and consistency of the entire RoPE operation.

\subsection{Softmax}

\begin{figure*}
	\centering
		\includegraphics[trim={0cm 0cm 0cm 0cm},clip,scale=.3]{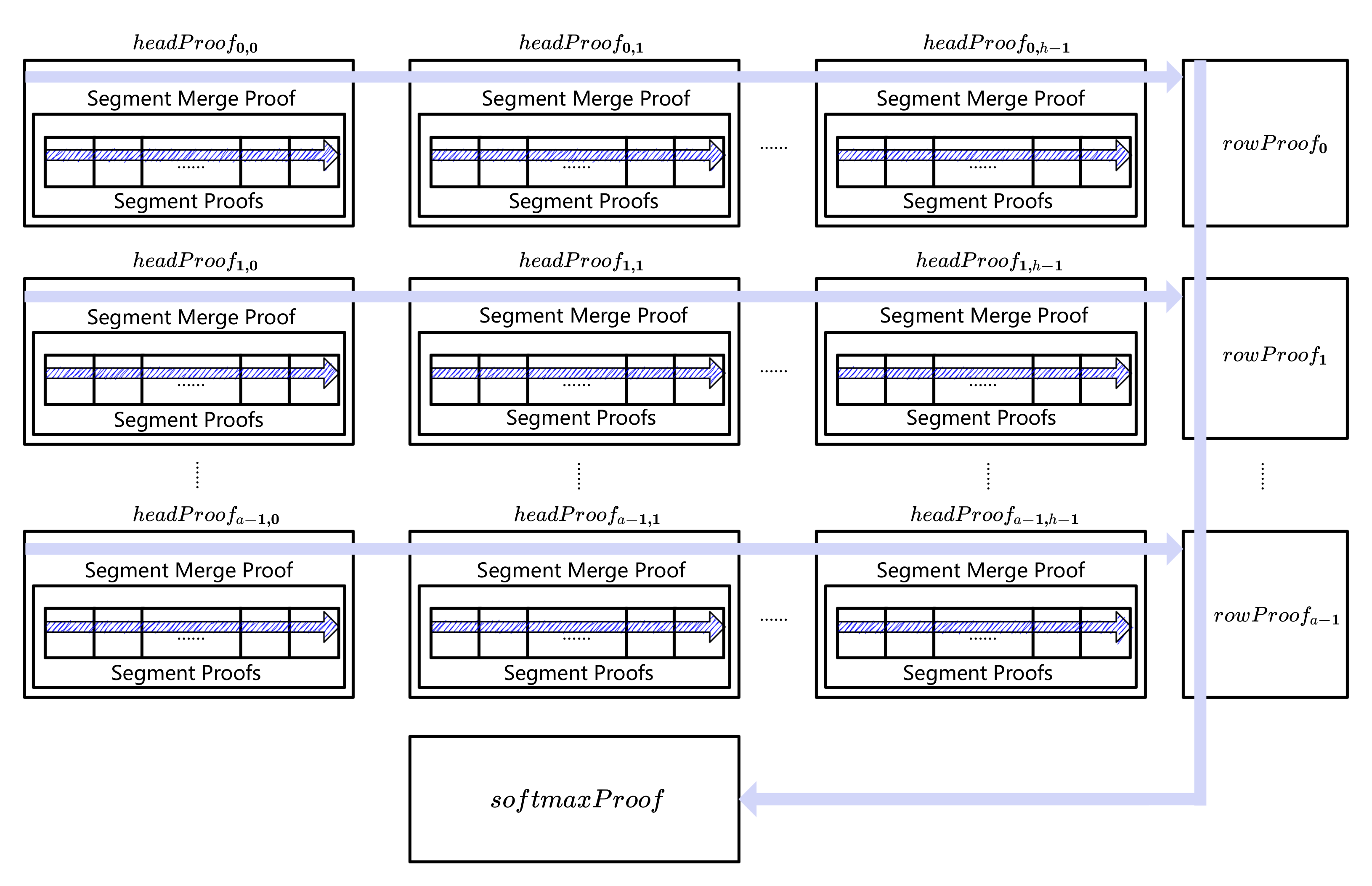}
	\caption{The proof composition of the Softmax component}
	\label{FIG:8}
\end{figure*}

Softmax is a normalization function that transforms a vector of real numbers into a probability distribution, amplifying larger values while suppressing smaller ones. It is defined as

$$\sigma(x_i) = \frac {e ^ { x_i } } { \sum _ { j } e ^ { x_j } }, $$

\noindent ensuring that every output is positive and that the entire vector sums to one. This makes Softmax an essential component in classification tasks and attention mechanisms, where probabilities must be well-behaved.

To integrate Softmax computation within our zero-knowledge framework, we construct a precise fixed-point, integer-based approximation. The input vector $X$ is first scaled by $2^q$, enabling accurate arithmetic inside the circuit. Using the identity $e^x = 2^{x \cdot log_2e}$, exponentiation is reduced to computing powers of two, where fractional exponents  $2^y$, for $0 \leq y < 1$, are retrieved from a precomputed lookup table. The procedure unfolds as follows:

1. Compute $LOG2E\_Q = round(log2(e) * 2^q)$ and build a lookup table $EXP2\_FRAC\_TABLE$ of length $2^l$ , where $EXP2\_FRAC\_TABLE[i] = round( 2^{-i/2^l} * 2^q )$ for $0 \leq i < 2^l$;

2. For numerical stability, compute $\Delta_i = x_i - x_{\max}$, where $x_{\max} = \max(x)$.

3. For each $i$, derive

$$y_i = \frac {\Delta_i \cdot LOG2E\_Q } {2^q}$$

and split it into integer and fractional parts:
$$
k_i = (-y_i) \gg q, \qquad f_i = (-y_i) \ \&\ (2^q - 1).
$$

4. Extract the high $l$ bits of $f_i$ as 

$$idx_i = f_i \gg (q - l),$$

\noindent query the lookup table to obtain $t_i = EXP2\_FRAC\_TABLE[idx_i]$, and compute the exponent approximation
$$
w_i = t_i \gg k_i.
$$

5. Normalize to obtain the Softmax probability:

$$p_i = \frac{w_i \cdot 2^q}{\sum_{j}(w_j)}$$.

To support modern multi-head attention mechanisms, we design a proof construction that efficiently verifies Softmax across multiple heads. Because the dimensionality of each head may grow dynamically during inference, we partition every head into fixed-length segments, padding unused entries with a large negative constant so that their Softmax outputs evaluate to zero. The verification process proceeds in five stages:

1. For each row, every head is divided into segments, and a SNARK proof is generated for each segment. During this process, we also accumulate the segment-wise sums of $w_i$ and the $ZMul$ values for both the input $X$ and output $Y$.

2. All segment proofs within a head are merged into a single proof that certifies the correctness of the entire head.

3. The merged head proof is wrapped while checking the aggregated sum of $w_i$, producing a head-level proof.

4. All head proofs in a row are merged into a row-level proof attesting to the correctness of the entire row’s Softmax computation.

5. Finally, all row proofs are recursively merged to produce a single proof that verifies the correctness and consistency of the entire multi-head Softmax operation.

The overall structure is illustrated in Figure 8.

\subsection{Sigmoid and Sigmoid Linear Unit(SiLU)}

The Sigmoid function is a classical nonlinear activation that maps any real-valued input to a smooth, bounded range between 0 and 1:

$$
\sigma ( x ) = \frac { 1 } { 1 + e ^ { - x } },
$$

\noindent turning raw inputs into probability-like values that vary smoothly with respect to the input.

The Sigmoid Linear Unit (SiLU), also known as Swish, refines this idea by multiplying the input with its sigmoid activation:

$$
SiLU(x) = x \cdot \sigma ( x ),
$$

\noindent offering smoother gradients and often superior optimization dynamics compared to traditional activations such as ReLU.

To incorporate these activations into our zero-knowledge proof system, we design a fully integer-based approximation that remains compatible with finite-field arithmetic. Following the same approach used for Softmax, we rely on the identity $e^{-x} = 2^{-x \cdot log_2e}$ and a precomputed lookup table for fractional powers of two. The process unfolds as follows:

1. Compute $LOG2E\_Q = round(log2(e) * 2^q)$ and generate a lookup table $EXP2\_FRAC\_TABLE$ of length $2^l$ , where $EXP2\_FRAC\_TABLE[i] = round( 2^{i/2^l} * 2^q )$ for $0 \leq i < 2^l$;

2. For each input $x_i$, calculate

$$y_i = \frac {-x_i \cdot LOG2E\_Q } {2^q}$$ and split it into integer and fractional parts:


$$
k_i = y_i \gg q, \qquad f_i = y_i \& \bigl((1 \ll q) - 1\bigr).
$$

3. Extract the top $l$ bits of the fractional component,

$$
idx_i = f_i \gg (q - l),
$$

\noindent and use it to query the lookup table:

$$
t_i = EXP2\_FRAC\_TABLE[idx_i].
$$

\noindent The approximate exponential is then

\[
u_i = 
\begin{cases}
t_i \gg (-k_i), & \text{if } k < 0, \\
t_i \ll k_i,  & \text{if } k \geq 0.
\end{cases}
\]

4. Using this approximation, we compute the sigmoid and SiLU activations:

$$
\sigma ( x_i ) = \frac { 2^{2q} } { 2^q + u_i },
$$ and $$
SiLU(x_i) = \frac { x_i \cdot \sigma ( x_i ) } { 2^q}.
$$

To support arbitrary model dimensions, we build a hierarchical SNARK proof system:

\begin{itemize}

\item The input matrix $X$ is divided into rows, and each row into fixed-size segments.

\item Each segment is proven independently, and the resulting proofs are merged into row-level proofs.

\item Finally, all row proofs are recursively combined into a single global proof certifying the correctness of the entire Sigmoid or SiLU computation.

\end{itemize}

The full proof architecture for Sigmoid and SiLU is illustrated in Figure 9.

\begin{figure*}
	\centering
		\includegraphics[trim={0cm 2cm 0cm 0cm},clip,scale=.3]{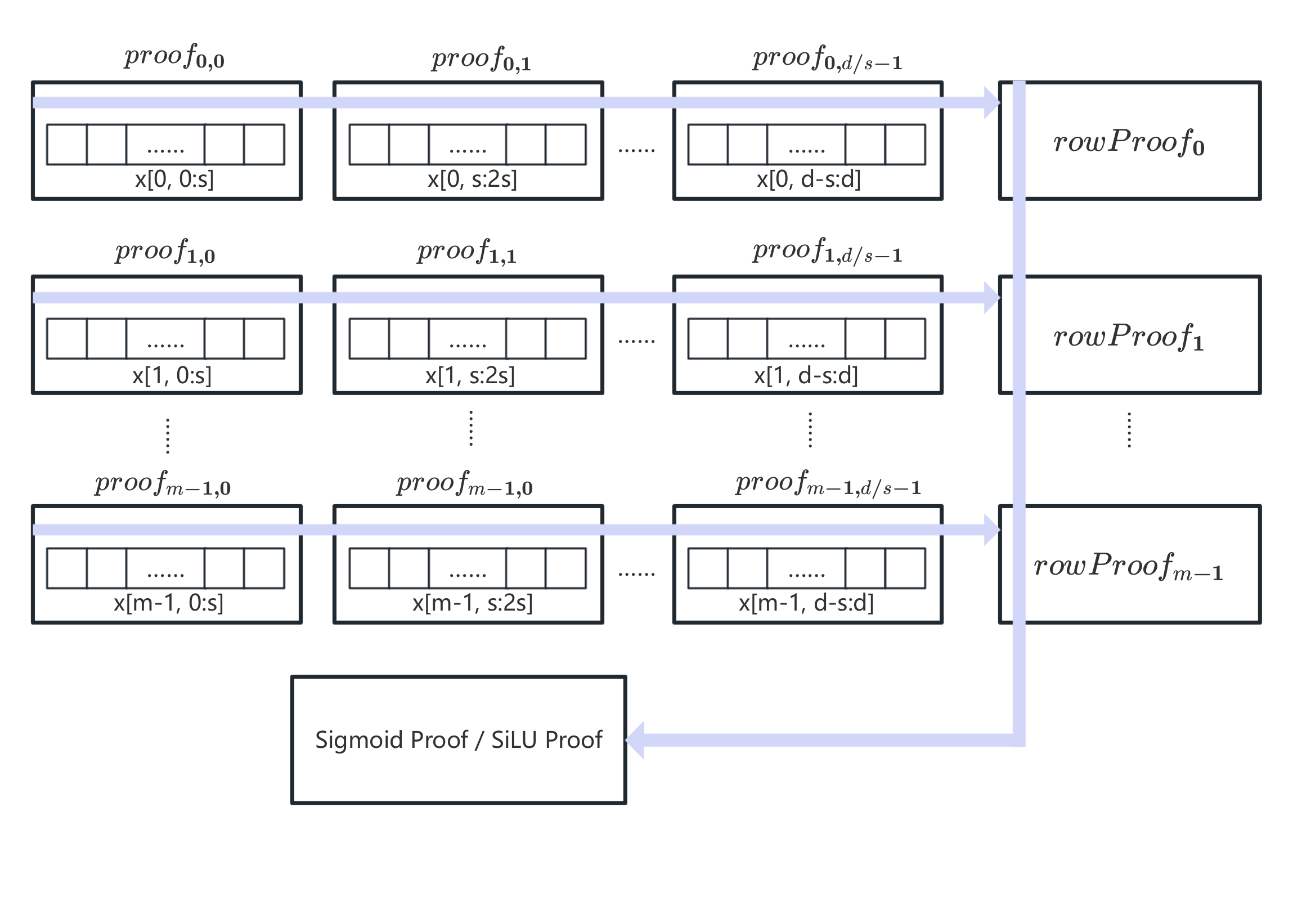}
	\caption{The proof composition of the Sigmoid and SiLU component}
	\label{FIG:9}
\end{figure*}

\subsection{Sort and top-k selection}

We design a zero-knowledge proof construction that supports both full sorting and top-$k$ selection. Suppose we have two lists $A$ and $B$, each of length $n$. We associate each list with a characteristic polynomial:

$$P_A(t) = \prod\limits_{i}(t - a_i), \qquad P_B(t) = \prod\limits_{i}(t - b_i).$$

By the Schwartz–Zippel Lemma, if $A$ and $B$ are \textbf{not} permutations of one another, then for a randomly chosen evaluation point $z$, the probability that
$$P_A(z) = P_A(z)$$

\noindent still holds is at most $\frac{\deg(P)}{|F|}$, which is negligible in a large field.

This observation forms the basis for verifying the correctness of a sorting operation in zero-knowledge. Given an original (unsorted) list $L$ and a candidate output $L'$, the prover demonstrates that:

\noindent 1. $P_L(z) = P_{L'}(z)$, ensuring that $L$ and $L'$ contain exactly the same multiset of values, and

\noindent 2. $L'$ is monotonically non-decreasing, i.e., $L'[i-1] \le L'[i]$ for all $1 \le i < n$.

\noindent When both conditions hold, the verifier can be confident that $L'$ is the correct sorted permutation of $L$. Furthermore, selecting the first $k$ elements of $L'$ immediately yields a verifiable top-$k$ result, proving that the chosen elements are indeed the $k$ smallest in the original list $L$.

\subsection{Element-wise Multiplication/Addition}

Element-wise operations fit naturally into our proof framework and can be verified with minimal overhead. Consider the example of element-wise addition $X + B = Y$, where $X$ and $B$ have identical shapes. We begin by splitting both matrices row-wise, and then further divide each row into fixed-size segments. Each segment is independently verified through a SNARK proof. The segment proofs are merged into row-level proofs, and these row proofs are recursively composed into a final proof that certifies the correctness of the full element-wise operation.

Throughout this procedure, we accumulate the $ZMul$ values of both $X$ and $Y$, ensuring consistent linkage with the outputs of previous components and the inputs to subsequent ones. If $B$ corresponds to a constant model parameter, we additionally accumulate its hash to guarantee parameter integrity. If, instead, $B$ is the output of an earlier computation, we also accumulate its $ZMul$ value to confirm that it indeed originates from the preceding step.

\section{ZK-DeepSeek, a SNARK-verifiable LLM}

\begin{figure*}
  \centering
    \includegraphics[trim={0cm 0cm 0cm 0cm},clip,scale=.3]{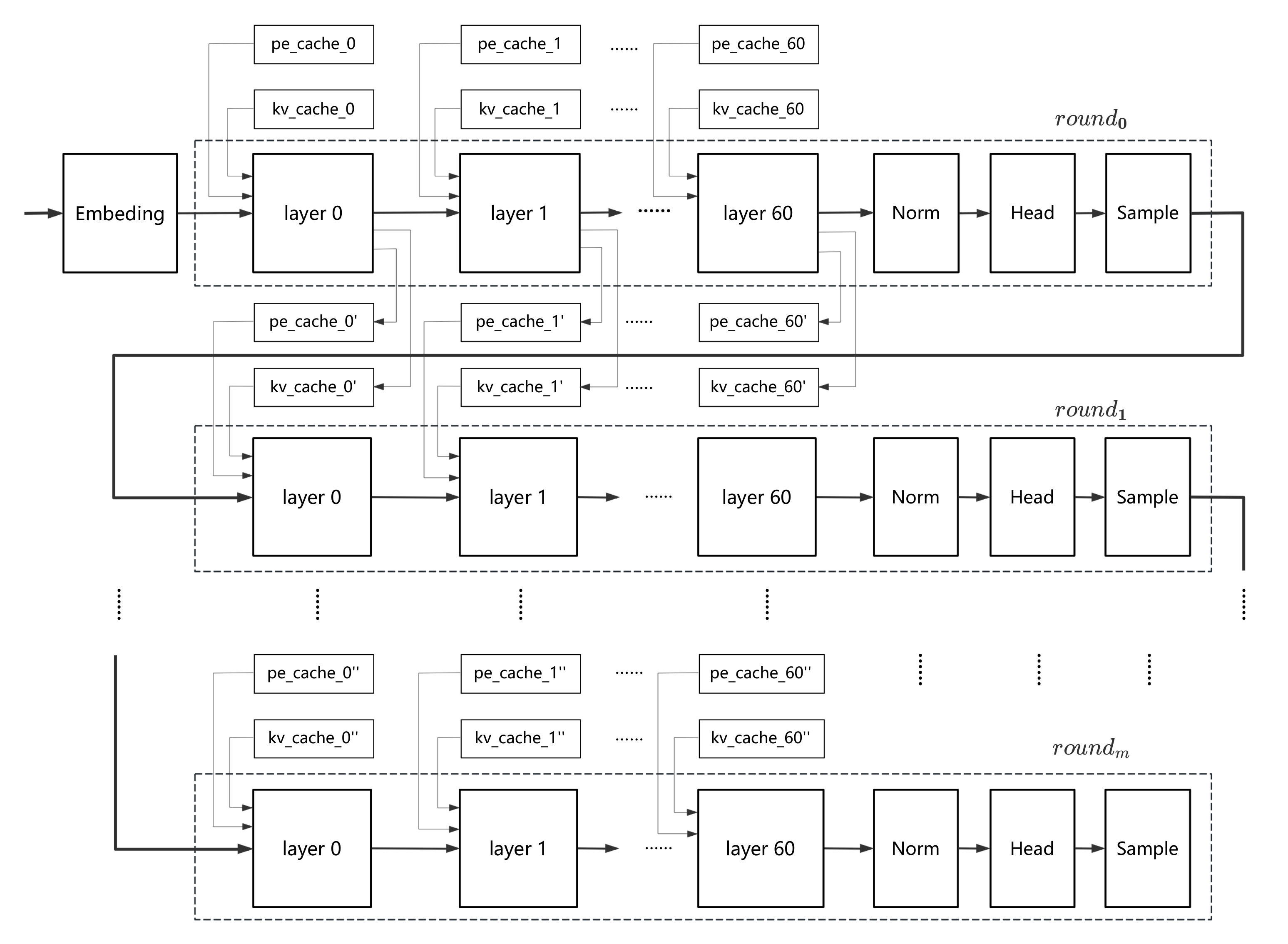}
  \caption{The inference pipeline of DeepSeek}
  \label{FIG:10}
\end{figure*}

DeepSeek is an advanced AI assistant designed to deliver intelligent, accurate, and helpful responses across a wide range of tasks, including general knowledge, coding, research, and writing. In this section, we focus on its largest variant,  \textbf{DeepSeek-V3}, using it as a representative example to demonstrate the capabilities of our zero-knowledge verification framework.

\subsection{DeepSeek Structure Overview}

DeepSeek-V3 follows a transformer-style inference pipeline composed of an embedding module, a stack of transformer layers, a normalization module, an output head, and a sampling module. Each layer integrates a Multi-Head Latent Attention (MLA) mechanism, a Mixture-of-Experts (MoE) feed-forward network, and the corresponding normalization components. The complete architecture is shown in Figure~\ref{FIG:1}.

As illustrated in Figure~\ref{FIG:1}, the input tokens are first mapped into vocabulary embeddings, which then pass through the stacked layers. Throughout inference, the model maintains two caches: the Key-Position-Rotated cache ($pe\_cache$) and the Key-Value cache ($kv\_cache$). After each layer’s computation, its output becomes the input to the next layer, and both caches are updated with newly derived values. This design enables efficient incremental inference and supports long-context reasoning which is an essential capability for large-scale language models.

\subsection{SNARK Verification of MLA}

\begin{figure}[htbp]
	\centering
		\includegraphics[trim={1cm 0cm 0.96cm 0cm},clip,scale=.28]{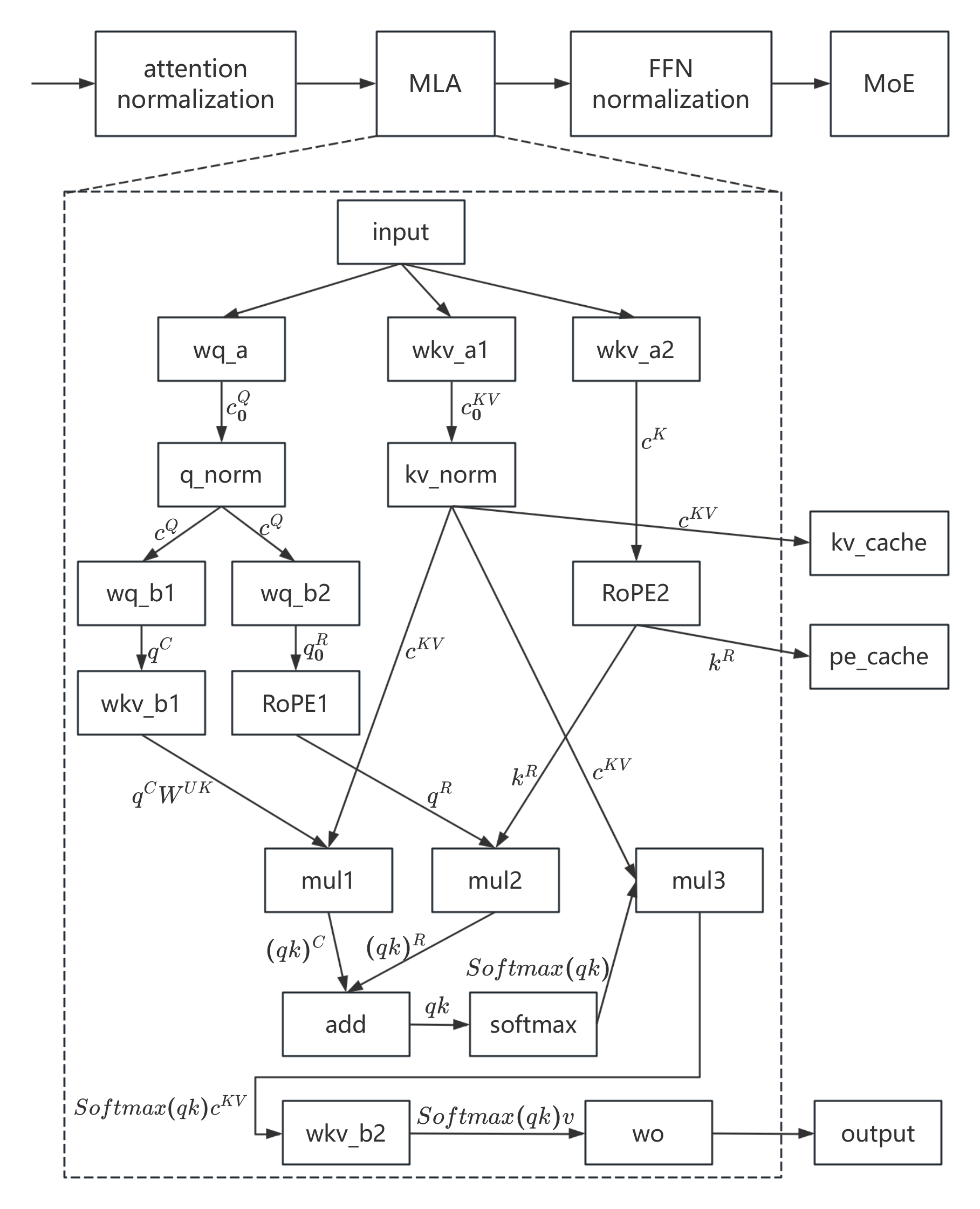}
	\caption{The dataflow of MLA}
	\label{FIG:11}
\end{figure}

To make MLA more suitable for SNARK verification, we introduce several small but important structural adjustments to the original design. Specifically, we split the $wq\_b$ tensor multiplication into two stages ($wq\_b1$ and $wq\_b2$) and likewise decompose the $wkv\_a$ multiplication into $wkv\_a1$ and $wkv\_a2$. These refinements allow us to express MLA as a sequence of simpler modular proof components that can be verified and composed more efficiently. The resulting dataflow is illustrated in Figure~11, and the functionality of each component is summarized below.

\begin{itemize}
    \item $wq\_a$: Down-projection matrix multiplication for queries, producing a compressed latent representation of the queries.
    \item $wkv\_a1$: Down-projection matrix multiplication for KV values, generating a compressed latent representation.
    \item $wkv\_a2$: Down-projection matrix multiplication for Key values, producing a compressed latent representation of the Key information.
    \item $q\_norm$: RMS normalization applied to the query vectors.
    \item $wq\_b1$: Up-projection matrix multiplication for queries, producing query representations \textbf{without} positional information.
    \item $wq\_b2$: Up-projection matrix multiplication for queries, producing query representations \textbf{with} positional information.
    \item $kv\_norm$: RMS normalization applied to KV data. The resulting $c^{KV}$ is stored in the KV cache $kv\_cache$.
    \item $wkv\_b1(W^{UK})$: Up-projection matrix multiplication for KV data, simultaneously extracting Key information from the combined Key-Value representation. In the original MLA design, $q \cdot k$ is optimized as

\begin{equation*}
\begin{split}
    q \cdot k = q \cdot (W^{UK} \cdot c^{KV}) \\
    = (q \cdot W^{UK}) \cdot c^{KV} \\
\end{split}
\end{equation*}
    
    \item $wkv\_b2(W^{UV})$: Up-projection matrix multiplication for Key-Value data, recovering Value vectors. Following MLA’s optimization,

\begin{equation*}
\begin{split}
    Softmax(qk) \cdot v = Softmax(qk) \cdot (c^{KV} \cdot W^{UV}) \\
    = ( Softmax(qk) \cdot c^{KV}) \cdot W^{UV} \\
\end{split}
\end{equation*}

    \item $RoPE1$: Rotary Positional Embedding applied to queries, injecting positional information into query vectors.
    \item $RoPE2$: Rotary Positional Embedding applied to Keys, injecting positional information into Key vectors. The result $k^R$ is cached in $pe\_cache$.
    \item $mul1$: Matrix multiplication between queries and Keys \textbf{without} positional information.
    \item $mul2$: Matrix multiplication between queries and Keys \textbf{with} positional information.
    \item $add$: Element-wise addition that combines the attention scores from the positional and non-positional branches.
    \item $mul3$: Matrix multiplication between attention scores 
    
    ($Softmax(qk)$) and the compressed KV vectors ($c^{KV}$).
    \item $wo$: Output projection matrix multiplication applied to the final attention outputs.
\end{itemize}

All MLA components map cleanly onto the SNARK building blocks described in Section~3, including GeMM, RMSNorm, RoPE, and Softmax. Table~\ref{tab:settings} provides a detailed correspondence. To generate a proof for the full MLA computation, we first build proofs for each individual component (e.g., $proof_{wq\_a}$, $proof_{q\_norm}$), then recursively merge them following the MLA computation graph. This incremental composition ultimately yields the unified proof $proof_{MLA}$, attesting to the correctness of the entire MLA process.

\begin{table}[htbp]
  \caption{The mapping between SNARK components and MLA operations}
  \label{tab:settings}
  \begin{tabular}{lp{5cm}}
    \toprule
     & Components \\
    \midrule
    GeMM & wq\_a, wkv\_a1, wkv\_a2, wq\_b1, wq\_b2, wkv\_b1, wkv\_b2, mul1, mul2, mul3, wo \\
    RMSNorm & q\_norm, kv\_norm \\
    RoPE & RoPE1, RoPE2 \\
    Softmax & softmax \\
    Element-wise Addition & add \\
    \bottomrule
  \end{tabular}
\end{table}

\subsection{SNARK Verification of MoE}

The Mixture-of-Experts (MoE) module consists of a gating network, a set of shared experts, and a large pool of regular experts. During inference, the shared experts are always active, while the regular experts are dynamically selected based on gating weights computed from the input. As shown in Figure 12, each chosen expert produces an output $y_i$, which is scaled by its corresponding gate weight in component $mul1$. These weighted expert outputs are then combined with the output of the shared experts $z$ to produce the final MoE result.

In Figure 12, the gate component and the Multi-Layer Perceptron (MLP) structures are depicted separately. Both shared experts and regular experts follow standard MLP architectures, which allows every part of the MoE module to be instantiated directly from the SNARK components introduced in Section 3. A detailed mapping between MoE operations and their SNARK building blocks is provided in Table 2.

\begin{figure}
	\centering
		\includegraphics[trim={0cm 0cm 0.9cm 0cm},clip,scale=.28]{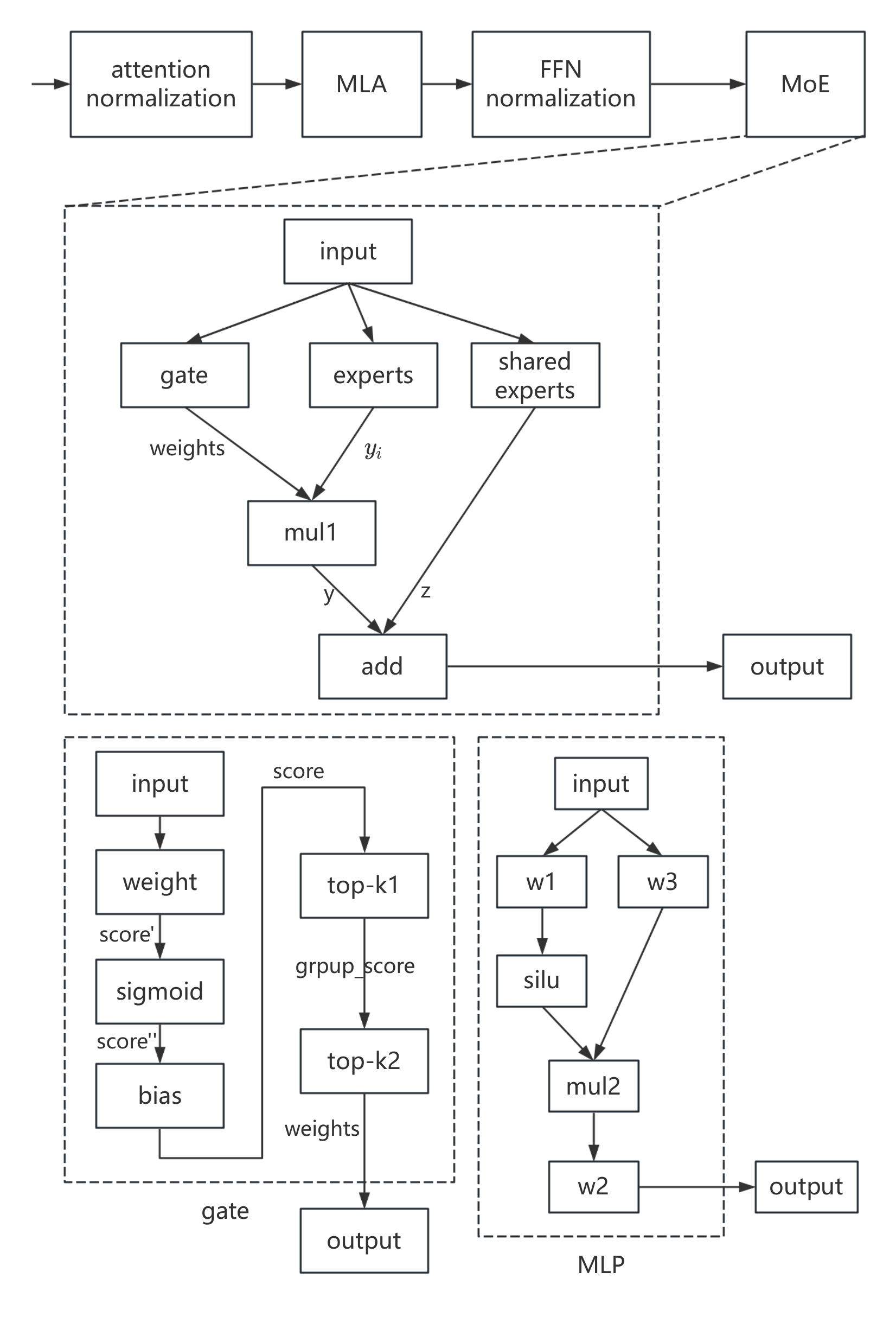}
	\caption{The dataflow of MoE}
	\label{FIG:12}
\end{figure}

\begin{table}[t]
  \caption{The mapping between SNARK components and MoE operations}
  \label{tab:settings2}
  \begin{tabular}{lp{4cm}}
    \toprule
     & Components \\
    \midrule
    GeMM & weight, w1, w2, w3 \\
    Sigmoid & sigmoid, kv\_norm \\
    Silu & silu \\
    Top-k & top-k1, top-k2 \\
    Element-wise Multiplication & mul1, mul2 \\
    Element-wise Addition & add, bias \\
    \bottomrule
  \end{tabular}
\end{table}
\section{Evaluation}
\label{sec:eval}

We implement \textbf{ZK-DeepSeek} based on the \textbf{DeepSeek-V3-0324} release \cite{deepseek-v3-0324}. To provide a representative and meaningful assessment of our framework, we benchmark several core components of the model, including embedding, the $wq\_a$ submodule of MLA, softmax, sigmoid, and attention normalization, covering both linear and nonlinear computations widely used in LLM inference.
Our full model and source code are open-sourced at:

\begin{itemize}
    \item https://huggingface.co/arcstar-lab/ZK-DeepSeek
    \item https://github.com/arcstar-lab/ZK-DeepSeek
\end{itemize}

\subsection{Experiment Setup}

\textbf{Hardware}
All experiments are performed on a high-end workstation equipped with an Intel i9-14900KF 32-core CPU, an NVIDIA GeForce RTX 5090 GPU, 64 GB of RAM, and 6 TB of SSD storage. Of this storage, 800 GB is allocated as virtual memory to accommodate the large intermediate states generated during model conversion. The workstation is connected to a 500-Mbps local network.

\medskip

\noindent\textbf{Software}
ZK-DeepSeek is built using Python 3.12 and CUDA 12.9 (driver 575.64.05). The SNARK verification backend is implemented in o1js 2.10 running on Node.js v24.8. Inference kernels are written in Python and CUDA, while all zero-knowledge circuits and proof generation logic are implemented in o1js.

\subsection{Experiment Implementation}

DeepSeek-V3 is originally trained and executed with floating-point formats, primarily BF16 with some FP32, and its weights are stored in BF8 and BF16. Floating-point arithmetic, however, is inefficient to verify within SNARKs. To address this, we quantize the entire model, both in storage and during execution. All linear and nonlinear computations are performed using integers: Int64 or Int32 for storage and Int64 for computation, with occasional Int128 intermediates to avoid overflow.

This quantization significantly increases model size: the full ZK-DeepSeek model expands to \textbf{~2.5 TB}, compared to the original DeepSeek-V3’s \textbf{680 GB}. Because the representable range of Int64 is narrower than that of BF16/FP32, we conduct statistical analyses of DeepSeek-V3’s weight distributions and approximate extremely rare values (< 0.01\% frequency) to preserve numerical stability.

The second major modification concerns model loading and inference. DeepSeek-V3 contains up to 671 billion parameters, which would require over 600 GB of GPU memory if loaded at once—far beyond typical hardware limits. To make inference tractable, we implement a \textbf{layer-by-layer execution strategy}. Each layer is stored separately on disk and is loaded into GPU memory only when needed. Experts in the MoE layers are also stored independently and loaded dynamically during inference.

During inference, we iteratively load only the necessary components:

\begin{itemize}
\item Load the parameters for the current layer along with the Key-Value cache ($kv\_cache$) and Key-Position-Rotated cache ($pe\_cache$).
\item Execute MLA attention for the current layer, update the caches, and write them back to SSD.
\item Run the MoE gating logic to determine the top-$k$ experts.
\item Dynamically load the selected experts and compute their outputs.
\item Combine the outputs of selected experts with those of the shared experts to produce the final layer output.
\end{itemize}

This design reduces GPU memory usage from over \textbf{600 GB} to \textbf{under 24 GB}, enabling both DeepSeek-V3 and ZK-DeepSeek to run on a single RTX 4090 or 5090 GPU. Although dynamic loading and cache transfers slow down inference, the approach remains fully compatible with our SNARK-verifiable architecture and makes large-scale experimentation feasible on consumer-grade hardware.

\subsection{Result and Discussion}

Table 3 summarizes the performance results of our evaluation.

The \textbf{embedding} component divides each row of the vocabulary embedding matrix into segments of size 224. After constructing and recursively merging all segment and row proofs, the total proving time for this component is \textbf{4,823 seconds}.

Both \textbf{attn\_norm} and \textbf{q\_norm} correspond to RMSNorm verification. Their inputs are partitioned by rows and then by segments, with segment sizes of 112 for attn\_norm and 48 for q\_norm. Comparing their row-level and full proof times, we observe that attn\_norm is significantly more efficient. This indicates that, when the resulting constraint size remains tractable, larger segments can noticeably improve proving efficiency.

The \textbf{RoPE1} component verifies the positional rotation applied to queries. Each row is split into 128 attention heads, each of dimension 64. Due to the computational complexity of RoPE, the total proving time is relatively large, reaching \textbf{19,275 seconds}.

The \textbf{softmax\_qk} component verifies the Softmax operation applied to the $qk$ matrix in MLA. The input is a (24 $\times$ 3072) matrix, where 3072 corresponds to the combined width of 128 heads (each 24 elements wide). Because the effective head width increases during inference, we construct the proof incrementally using segments of size 32. Similar to RoPE, the full proving time also reaches \textbf{39,456 seconds}, reflecting the inherent complexity of Softmax computation.

The \textbf{sigmoid\_gate} component verifies the sigmoid transformation applied to MoE gating scores. Each row contains 256 elements (one per expert), and is further divided into segments of size 16. The total proving time for this component is \textbf{2,874 seconds}.

The \textbf{experts\_selector} component verifies the expert-selection logic in MoE. Each row again contains 256 experts, grouped into eight groups of 32. The system first selects four candidate groups using the top two experts per group, and then selects eight final experts from these candidates, requiring two rounds of top-$k$ selection. The full proving time for this component is \textbf{2,447 seconds}.

Finally, the \textbf{wkv\_a1} component verifies the matrix multiplication
$$
A^{a \times n} \cdot B^{n \times b}, \quad \text{where } a=24, n=7168, b=512.
$$
Following DeepSeek’s design, we use $B_2^{b \times n} = (B^{n \times b})^\top$ for computation. Both $A$ and $B_2$ are split by rows of length 7168 and further partitioned into segments of size 112. From $A$, we generate 64 $XProofs$; from $B_2$, 64 $WProofs$. These pairs are combined to produce 64 $XWProofs$, which are then recursively merged into a final proof certifying the correctness of the entire multiplication. The total proving time is \textbf{204,138 seconds}.

\begin{table*}[t]
  \caption{The metrics of different components}
  \label{tab:settings3}
  \begin{tabular}{ccccc>{\centering\arraybackslash}p{2cm}>{\centering\arraybackslash}p{1cm}>{\centering\arraybackslash}p{1cm}}
    \toprule
     Component & Type & Config & Proof type & Count & Avg proving time(s) & Proof size(K) & Verifying time(ms)\\
    \midrule
    \multirow{3}{*}{$embedding$} & \multirow{3}{*}{embedding} & \multirow{3}{3.8cm}{rows=24, Dim=7168, segmentDim=224} & segment & 768 & 2.44 & 32-36 & 326.73 \\ \cline{4-8}
            & & & row & 24 & 194.71 & 32-36 & 319.21 \\ \cline{4-8}
            & & & component & 1 & 4823 & 32-36 & 342.14 \\ \hline

    \multirow{3}{*}{$attn\_norm$} & \multirow{3}{*}{RMSNorm} & \multirow{3}{3.8cm}{rows=24, Dim=7168, segmentDim=112} & segment & 1536 & 2.67 & 32-36 & 324.55 \\ \cline{4-8}
            & & & row & 24 & 401.21 & 32-36 & 354.31 \\ \cline{4-8}
            & & & component & 1 & 9941 & 32-36 & 343.00 \\ \hline

    \multirow{3}{*}{$q\_norm$} & \multirow{3}{*}{RMSNorm} & \multirow{3}{3.8cm}{rows=24, Dim=1536, segmentDim=48} & segment & 768 & 2.68 & 32-36 & 320.88 \\ \cline{4-8}
            & & & row & 24 & 195.04 & 32-36 & 374.08 \\ \cline{4-8}
            & & & component & 1 & 4970 & 32-36 & 348.47 \\ \hline

    \multirow{3}{*}{RoPE1($q\_pe$)} & \multirow{3}{*}{RoPE} & \multirow{3}{3.8cm}{rows=24, Dim=8192, headDim=64, headCount=128} & head & 3072 & 2.61 & 32-36 & 355.38 \\ \cline{4-8}
            & & & row & 24 & 791.58 & 32-36 & 353.52 \\ \cline{4-8}
            & & & component & 1 & 19275 & 32-36 & 333.49 \\ \hline

    \multirow{3}{*}{$softmax\_qk$} & \multirow{3}{*}{Softmax} & \multirow{3}{3.8cm}{rows=24, Dim=3072, headDim=24, headCount=128, segmentDim=32} & head & 3072 & 8.97 & 32-36 & 324.87 \\ \cline{4-8}
            & & & row & 24 & 1632.25 & 32-36 & 319.67 \\ \cline{4-8}
            & & & component & 1 & 39456 & 32-36 & 362.61 \\ \hline

    \multirow{3}{*}{$sigmoid\_gate$} & \multirow{3}{*}{Sigmoid} & \multirow{3}{3.8cm}{rows=24, Dim=256, segmentDim=16} & segment & 384 & 3.46 & 32-36 & 353.51 \\ \cline{4-8}
            & & & row & 24 & 107.63 & 32-36 & 384.87 \\ \cline{4-8}
            & & & component & 1 & 2874 & 32-36 & 334.08 \\ \hline

    \multirow{6}{*}{$experts\_selector$} & \multirow{6}{*}{top-$k$} & \multirow{6}{3.8cm}{rows=24, Dim=256, groupDim=32, groupCount=8} & group & 192 & 2.44 & 32-36 & 321.78 \\ \cline{4-8}
            & & & groupRow & 24 & 44.33 & 32-36 & 365.04 \\ \cline{4-8}
            & & & sortedGroup & 192 & 2.95 & 32-36 & 362.55 \\ \cline{4-8}
            & & & sortedGroupRow & 24 & 48.75 & 32-36 & 341.00 \\ \cline{4-8}
            & & & component & 1 & 2447 & 32-36 & 357.92 \\ \hline

    \multirow{3}{*}{$wkv\_a1$} & \multirow{3}{*}{GeMM} & \multirow{3}{3.8cm}{rows(a)=24, inDim(n)=7168, outDim(b)=512, segmentDim=112} & XProof & 64 & 130.72 & 32-36 & 374.72 \\ \cline{4-8}
            & & & WProof & 64 & 3045.80 & 32-36 & 335.26 \\ \cline{4-8}
            & & & component & 1 & 204138 & 32-36 & 322.61 \\

    \bottomrule
  \end{tabular}
\end{table*}
\section{Conclusion}
\label{sec:conclusion}

We have introduced a SNARK framework specifically designed for neural networks, capable of handling models at extremely large scales. To demonstrate its practicality, we applied the framework to DeepSeek-V3 and constructed ZK-DeepSeek, a fully SNARK-verifiable large language model. While verifying an LLM with SNARKs remains computationally intensive (especially in our current CPU-based implementation), our results clearly show that verifiable AI through zero-knowledge proofs is not only possible but already within reach.

Our next priority is to develop a GPU-accelerated version of the framework, which we anticipate will yield performance improvements by several orders of magnitude and make verifiable AI substantially more practical. We are actively seeking funding and collaborators to help advance this line of research. If you are interested in supporting or partnering with us, we warmly invite you to get in touch by email.


\bibliographystyle{plainnat}
\bibliography{bib}

\appendix


\end{document}